# FRAGMENTARY ELECTRODYNAMICS OF SOLITONS EMITTED BY ATOMS

Pavel S. Kamenov

Sofia University, Faculty of Physics

**Summary**

In the recent years there was published some papers in which the photons are represented as electromagnetic solitons [1,2,3]. All particles – solitons – represent some electromagnetic field restricted in a very small volume, length, cross-section and propagate in vacuum with light velocity in one and the same directions at a very great unrestricted distances [3]. These unordinary properties of the electromagnetic field – soliton – require some more detailed investigations of the dynamics of interactions of charged particles and electromagnetic fields.

In this paper I make an attempt to solve in part (fragmentary) some questions about energy density of the soliton, electrodynamics of soliton and the simplest hydrogen atom, the acceleration of a charged particle, the path of the electron wile it changes its position, the emitted electromagnetic soliton energy and the electron in a stationary state. The descriptions are restricted only to the properties of the solitons in paper [3] and to the hydrogen atom, but I think that obtained results could be applied to the more complicated systems.

**Introduction**.

As was shown in [3], the electromagnetic field of a single photon must be concentrated in a very small volume, V. The relation between the maximal electric field ($E_0$) of a soliton and the frequency ($\omega$) of a free photon is: $E_0^2 = (8m_0\hbar/e^2)\omega^3$, where $\omega$ is the frequency obtained from interference phenomena, de Broglie's frequency; $m_0$ and e are the mass and electric charge of the electron. The energy of the photon is $\hbar\omega = E_0^2 V$; volume $V = (e^2/8m_0\hbar\omega^2)$; the energy density of the soliton is $\rho_s = E_0^2$; the effective time of action is $t_e = 1/2\omega$ and the effective length ($l_e$) of the electromagnetic field of the soliton and the wave length ($\lambda$) of de Broglie are related:

$l_e = \lambda/4\pi$. So, the electromagnetic field occupies only a small part of $\lambda$ and a very small part of the photon wave package (wave function).

### *1. Consistence with Planck's density of radiation.*

In the Fig.1 schematically is presented a soliton with the above mentioned properties. $S_e = e^2/8m_0c\omega$ is macroscopic cross-section of the soliton.

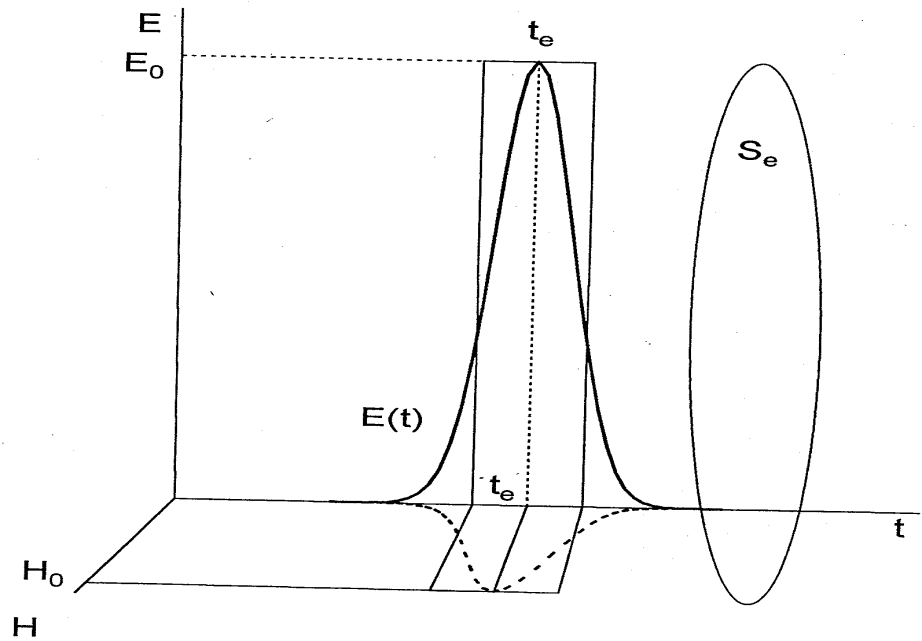

**Fig.1.** *Schematic representation of soliton electromagnetic field. The shape of the field is not known exactly, but the effective time is $t_e = 1/2\omega$.*

As it is known, Planck's density of radiation is

$$\rho(\omega) = E^2 = \frac{\hbar\omega^3}{\pi^2 c^3} \frac{d\omega}{[\exp(\hbar\omega/kT) - 1]} \quad (1)$$

The unit frequency interval is $d\omega = 1 \text{ s}^{-1}$. The part which depends on temperature ($T$) is usually interpreted as the average number of photons ($\overline{N}$) in unit volume. So, (1) can be represented as

$$\rho(\omega) = E^2 = \frac{\hbar\omega^3}{\pi^2 c^3} \overline{N} \quad (2)$$

When $\bar{N} = 1$, equation (2) can be compared with the equation for the soliton energy density ($\rho_s$):

$$\rho_s = E_0^2 = \frac{8m_0\hbar}{e^2}\omega^3 \qquad (3)$$

Or

$$\frac{E^2}{E_0^2} = \frac{e^2}{8\pi^2 m_0 c^3} \approx 1.2 \times 10^{-25} \qquad (3a)$$

$$E = \sqrt{\frac{e^2}{8\pi^2 m_0 c^3}} E_0 \approx 3.4 \times 10^{-13} E_0$$

This confirms the assumption that in a larger volume in comparison with the real volume of the electromagnetic field, the energy density (larger volume) appears to be smaller [3]. The soliton energy density is consistent with energy density of Planck. The only difference between (2) and (3) is that the soliton volume (*V*) is different from the unit volume. One can calculate the average number ($\bar{N}$) of photons for which Planck's energy density (2) is equal to the energy density of a soliton (3).

$$\frac{E_0^2}{E^2} = 1 = \frac{K_0 \omega^3 \pi^2 c^3}{\hbar \omega^3 \bar{N}} \qquad (4)$$

For $K_0 = \frac{8m_0\hbar}{e^2}$ this number is a constant:

$$\bar{N} = \frac{K_0 \pi^2 c^3}{\hbar} = \frac{8\pi^2 m_0 c^3}{e^2} \approx 8.1 \times 10^{24} \qquad (5)$$

(5) is reciprocal to (3a). These results explain the first unsuccessful attempt to describe the photoelectric effect within the framework of the electromagnetic wave theory. It is clear that the electric field *E* does not change with time *t* and frequency ω, as it was thought before, but the number of photons in unit volume remains proportional to $E^2$. The soliton (particle) and the energy density is concentrated in a very small volume. The frequency ω is the frequency of de Broglie's field *D*, like for all particles. The amplitude *D* of the real field of de Broglie accounts for all interference phenomena and for all particles.

*2. Soliton electrodynamics and hydrogen atom*

As it is known from electrodynamics only when a charged particle is accelerated it can emit an electromagnetic field. In the hydrogen atom of Bohr the electron in a stationary state dose not emit the photons. The photon is emitted only when the radial coordinates change. In this time the electron changes the velocity passing from one upper excited state ($k$) to some lower state ($n$). If one know the time of transition, then acceleration of electron could be found. Up to now this time of transition is not known and something more, we accept that it is *useless* to think about this time.

On the other hand, if the photon can be represented as an electromagnetic particle – soliton – with a determined volume, length and cross-section [3], then effective action time of the soliton depends of it de Broglie's frequency ($\omega$) as $t_e = 1/2\omega$. It must be accepted that effective action time of a soliton electric field (emitted from hydrogen) corresponds to the effective time of spontaneous transition in hydrogen atom. One can find the acceleration of electron while the transition occur and knowing this acceleration it is possible to obtain the energy of the soliton (the energy of the emitted photon).

**Acceleration**. As was mentioned the effective time ($t_e = 1/2\omega$) of the soliton require a corresponding acceleration time of electron when the electron passes from upper ($k$) to lower level ($n$) in hydrogen atom. On the Fig.2 schematically are shown the velocities ($V_k$, $V_n$) and acceleration ($a$).

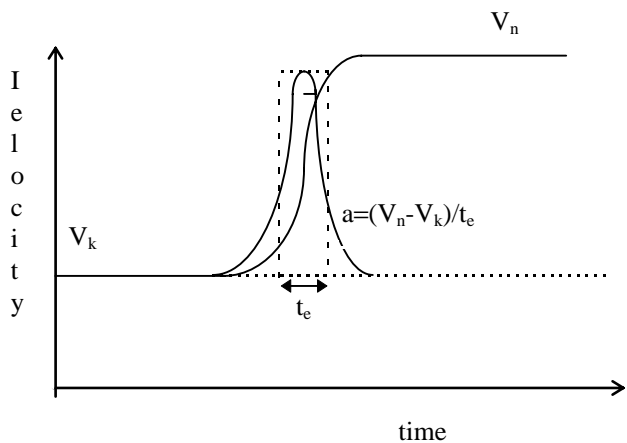

**Fig.2**. *A scheme of velocities ($V_k$, $V_n$) and acceleration ($a$). The shape of acceleration curve is not exactly known, but the effective time of velocity change is $t_e = 1/2\omega$.*

The effective time of acceleration and the shape of acceleration curve must correspond to the effective time of the soliton and to the shape of electric field of the soliton (Fig.1 and Fig.2). When the electron is in a stationary state (no acceleration) the electric field of the soliton is zero [4,5]. Knowing the time ($t_e$) and the velocities ($V_k$ and $V_n$) we can obtain the effective acceleration ($a$) of the electron when passing from upper level ($k$) to lower level ($n$). The corresponding velocities are:

$$V_k = e^2/\hbar k; \quad V_n = e^2/\hbar n \tag{6}$$

The effective time of acceleration correspond to the effective time of the emitted electromagnetic soliton ($t_e$) and effective acceleration must be:

$$a = (V_n - V_k)/t_e = (e^2/\hbar)(1/n - 1/k)/t_e \tag{7}$$

Substituting ($t_e = 1/2\omega$) we obtain:

$$a = 2\omega(V_n - V_k) = (2e^2\omega/\hbar)(1/n - 1/k) \tag{8}$$

The effective way (*path*) of electron ($H_{nk}$) while changing it state is:

$$H_{nk} = (a/2)t_e^2 + V_k t_e \tag{9}$$

Substituting here ($t_e$), (8) and $V_k$ from (6), one can find the effective path:

$$H_{nk} = (e^2\omega/\hbar)(1/n - 1/k)(1/4\omega^2) + e^2/\hbar k(1/2\omega)$$
$$H_{nk} = e^2/\hbar(1/2\omega)((1/2)(1/n - 1/k) + 1/k) \tag{10}$$

**Energy of the soliton**. According to the electrodynamics when the energy is emitted the average force (taking into account the force of reaction) is $F = am_0$. This force is a result of the Coulomb field of hydrogen atom and the reaction of electromagnetic field of the soliton. So, the energy ($E_{nk}$) which the accelerated electron emits can be found, substituting the necessary quantities in equation:

$$E_{nk} = FH_{nk} = am_0 H_{nk} \tag{11}$$

Substituting here (8) and (10) we obtain:

$$E_{nk} = \frac{e^4 m_0}{2\hbar^2}\left[\left(\frac{1}{n}-\frac{1}{k}\right)^2 + 2\left(\frac{1}{n}-\frac{1}{k}\right)\frac{1}{k}\right]$$

Or which is the same

$$E_{nk} = \frac{e^4 m_0}{2\hbar^2}\left[\left(\frac{1}{n^2}-\frac{1}{k^2}\right)\right] \qquad (12)$$

This is the energy which carry the electromagnetic particle – soliton – and the frequency of de Broglie (ω), as for all frequencies of the particles is:

$$E_{nk}/\hbar = \omega = \frac{e^4 m_0}{2\hbar^3}\left[\left(\frac{1}{n^2}-\frac{1}{k^2}\right)\right] \qquad (13)$$

As it is seen, in this way obtained energy (accelerated electron), coincide exactly with the results of Bohr and electromagnetic energy of the soliton, when it is emitted from a hydrogen atom [4].

From classical electrodynamics we know that a free electromagnetic field is proportional to the acceleration of charged particle, but the energy of the emitted field is redistributed in whole space and diminish with distance (r) as $1/r^2$. The field is maximal at perpendicular direction to the acceleration vector and is delayed in time as (t – r/c). Up to now it was not possible to calculate the effective acceleration of electron in hydrogen atom because the time of velocity change was not accessible for investigation. Even the questions "how long is this time", and "when the transition occur" are forbidden [4,5]. Now, the properties of the soliton determine this time and the effective acceleration is calculated. The energy of the emitted soliton is exactly equal to the energy losses in the hydrogen. Something more, all energy losses from the atom can be transferred (in vacuum) at a very great unrestricted distance as an electromagnetic particle – soliton. This means that whole electromagnetic energy is emitted in some well defined direction and the momentum of soliton is equal to the momentum of atom, but in opposite direction, as predicted by Einstein [6,7].

These properties of the photon-soliton and the hydrogen atom are not trivial and they must be examined in more details. If the solitons with these properties [3] exist in the nature, then the transitions in hydrogen atoms must take a time $t_e = 1/2\omega$ and the atom in a stationary state must be comparatively stable. In the following paragraphs I describe the hydrogen atom as a solitary (single) quantum system because every soliton must be emitted by only one solitary hydrogen atom.

*3. "When the transitions occur?"*

We must accept that the hydrogen atom emits a soliton only wile the electron and proton change the velocities. (Remember that the proton and electron move about the center of masses and change simultaneously its states). But why the electron cannot be accelerated when moving in a stationary state?

Contemporary quantum physics deals with statistical ensembles of quantum objects: atoms, nuclei, photons, electrons and other particles. The Bohr's model concern a solitary Hydrogen atom. This paper deals (also) with the single (separate) quantum object: one particle, one electron, one soliton, one single hydrogen atom, one nucleus, the "solitary quantum system" (SQS). Some specific properties of a SQS (hydrogen) are derived on the basis of experimental facts and the theory of contemporary quantum physics (QP). Thus, the properties of a solitary hydrogen atom do not contradict the results of QP, but allow us to think about and search for unknown and unexpected applications of quantum physics.

Remember that all quantum laws were initially derived from the results of experiments of statistical ensembles of quantum systems. Subsequently these laws were applied to solitary quantum systems which are the elementary constituents of the statistical ensemble. This is easy and trivial. Easy because it is not necessary to search for other properties of the solitary object and trivial because this transition does not contradict the laws which govern an ensemble of identical objects (quantum systems, QS). For a statistical ensemble of quantum systems the introduction of probabilities and the statistical interpretation of results are inevitable, but it is not sure that a solitary quantum system must be governed by the same principles. To be more specific, I can explain the above assertions with the help of an example:

***The law of radioactive decay,*** $N=N_0\exp(-t/\tau)$, was at first observed experimentally and after that derived from statistical considerations. N is the number of QS which have not decayed for the time t; $N_0$ is the number of QS at the initial moment of time (t = 0) and $\tau$ is the mean life time of all QS. This law concerns all decays of any excited states of nuclei, atoms, molecules and so on (except some deviations in very short and very long times). It is easy to transfer this law from a statistical ensemble to one solitary object by introducing the probability (W) that this object does not decay for a time (t): $W=N/N_0= \exp(-t/\tau)$. But this probability is a trivial application of a law which concerns *only a statistical ensemble of quantum objects*. This probability is not a proof that a solitary object does not have another cause for decay.

In the paper [8] it was shown that in the case of waves on the surface of a liquid the floating classical particles which pass through only one of the two opened slits are guided by the interfering surface waves in the same directions (angle θ) as predicted by quantum laws $|\Psi(\theta)|^2$=max; (the directions $|\Psi(\theta)|^2$=0 are not allowed.) This is an indirect confirmation of de Broglie's ideas that *wave and particle exist simultaneously and that this coexistence is real* [9]. Most of the scientists think that the field of de Broglie is not real and they accept the statistical interpretation of Born [10]. *One of the most often stressed disadvantages of the model of Bohr is the impossibility to determine (calculate) the probabilities of transition (intensities) of the emitted hydrogen lines...*

## 4. Return to the real unitary field-particle of de Broglie and to Bohr's model of hydrogen atom.

Why is the assumption that a wave-particle cannot exist simultaneously more real than de Broglie's assumption that they ***always exist simultaneously***?. The results of this work will show there is *simultaneous existence* of de Broglie's field and Bohr's atom ... ***and that (for one atom) no statistical interpretation is necessary***. De Broglie's waves in the hydrogen atoms are such that in the stationary state the mass of the electron ($m_0$), its velocity $V_n$, and average radius $r_n$ are related with the principal quantum number (*n*= 1,2,3....) according to:

$$m_0 V_n r_n = n\frac{h}{2\pi} = n\hbar \tag{14}$$

and the field-particle (electron) is in a potential well which keeps the electron in orbit *n*, and the electron cannot be accelerated (does not emit a photon). The length of de Broglie's wave $\lambda_n$ exactly satisfies the condition:

$$2\pi r_n = \frac{nh}{m_0 V_n} = n\lambda_n \tag{15}$$

De Broglie's unitary "wave-particle" is in a stationary ("steady") state which never changes. The "wave-particle" electron is bound together with the "wave-particle" proton by electromagnetic forces and de Broglie's real field. They interfere and remain in their potential well (position) forever, like classical particles on the surface of a liquid [8]. The field of de Broglie *is real and strong so that the electromagnetic forces cannot destroy this interference* and field-particle (electron) cannot be accelerated. To explain the decay of a stationary state it is necessary to assume some infinitely small "external perturbation" which would disturb the exact equalities (14) and (15) and (after some time of destructive interference) permit the transition to lower states (acceleration of electron). Only the state *n*=1 cannot be disturbed by an "infinitely small perturbation" because the field-particle cannot be destroyed (*n* cannot be smaller than 1 and $2\pi r_1 = \lambda_1$). In this case only if the perturbation energy is *sufficiently* great, can the electron make a transition from ground to upper levels by absorption, [8]).

I suppose that energetically excited the electron can randomly occur at any distance ($r_{ni}$) *around* the exact radius of the stationary orbit ($2\pi r_{ni} \approx n\lambda_n$ ;). The difference between the trajectory of the electron ($2\pi r_{ni}$) and $n\lambda_n$ can be very small, yet - destructive interference leads (after some time) to a transition to lower states. Imagine the "wave-particle" electron self-interferes as long as the minima of the wave coincides with the maxima of the preceding waves so that the amplitude (D) of the interfering electron-wave becomes $|D(t)|^2 = 0$. A transition (acceleration) occurs and energy is emitted. The greater the difference $|r_{ni} - r_n|$, the smaller the time necessary for destructive interference. If $|r_{ni} - r_n| \to 0$, the time for destructive interference would be very long [11]. When the energy of excitation is exact ($r_{ni} = r_n$), a true stationary

state would be established and without external perturbation this state could not be changed. So, it is evident that the wave-particle electron can be excited so, as to occur at all possible distances (r) from the proton.

## 5. Own lifetime of a single hydrogen atom.

In Fig.3 a schematic wave-particle in some excited state of the hydrogen atom is shown. The particle-wave electron moves from left to right (for example, $n = 2$). In Fig.3a) the velocity of the electron $V_n$ is such that $\lambda_n$ and $r_n$ correspond exactly to Bohr's conditions:

$$\lambda_n = \frac{h}{m_0 V_n} \qquad (16)$$

Such a wave-particle electron returns from the left always with the same phase and reiterates its motion for an infinitely long time. If the velocity of the electron ($V$) is slightly different, the new $\lambda$ will also be slightly different (compared with $\lambda_n$):

$$\lambda = \frac{h}{m_0 V} \qquad (17)$$

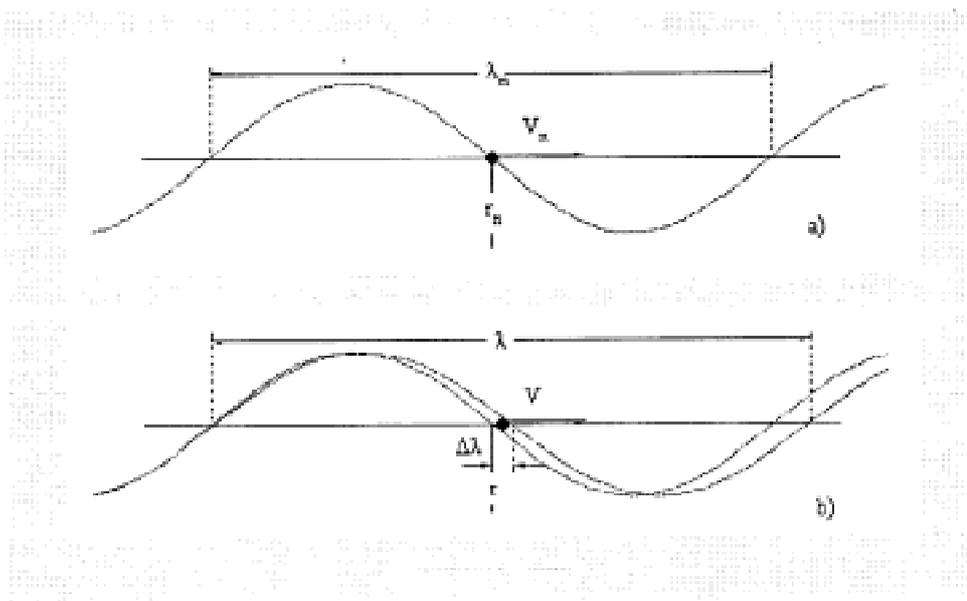

**Fig.3.** *A scheme of the first hydrogen excited states. Wave-particle electron and its interference; a) true stationary state; b) almost stationary state.*

Such a particle-wave electron would arrive from the left (Fig. 3b) with a slightly different phase. With time this difference increases and the moment when the sum of the amplitudes becomes zero (for the first time) can be calculated; electron is no more in the potential well of the wave (like classical particles, [8], when $|\Psi|^2=0$) and could be accelerated. When this occurs defines the time of life of this excited atom. The sum of the amplitudes of de Broglie' field (D) can be written (like classical particles, [8]):

$$D = \sin\left(\frac{2\pi}{\lambda}(Vt - r)\right) + \sin\left(\frac{2\pi}{\lambda}(Vt)\right) \tag{18}$$

where $r$ is the new radius which is only slightly different from $r_n$. The relation between $\lambda$, $\omega$ and velocity ($V$) is:

$$\lambda = \frac{2\pi V}{\omega} \tag{19}$$

Substitute this in (18) to give:

$$D = \sin\left(\omega t - \frac{\omega r}{V}\right) + \sin(\omega t) \tag{20}$$

In Bohr's model, $r/V = 1/\omega$, therefore (20) becomes

$$D = \sin(\omega t - 1) + \sin(\omega t) \tag{21}$$

which is the sum of the de Broglie's amplitudes (D), expressed by the time and the frequency of a not exactly stationary state. From (16) and (17) the small difference $\Delta\lambda$ and $\Delta\omega$ are found:

$$\frac{\Delta\lambda}{\lambda} = \frac{\lambda_n - \lambda}{\lambda} = \left(\frac{V}{V_n} - 1\right) = \frac{\omega}{\Delta\omega} \tag{22}$$

Taking into account that in Bohr's model

$$\frac{V}{V_n} = \sqrt[3]{\frac{\omega}{\omega_n}} = \sqrt[3]{\frac{\omega_n + \Delta\omega}{\omega_n}} \tag{23}$$

from (22) ($\omega$) is found:

$$\omega = \left(\Delta\omega\left(\sqrt[3]{\frac{\omega_n + \Delta\omega}{\omega_n}} - 1\right)\right) \tag{24}$$

The moment (t) when $|D|^2 = 0$ has to be found (the electron is not in the potential well of its wave and it is accelerated):

$$|D|^2 = |\sin(\omega t - 1) + \sin(\omega t)|^2 = 0 \qquad (25)$$

Hence

$$\sin(\omega t - 1) = -\sin(\omega t) \qquad (26)$$

or

$$\omega t - 1 = -\omega t; \qquad t = \frac{1}{2\omega} \qquad (26a)$$

So, substituting ω from (24), gives the necessary "own lifetime" (t):

$$t = \frac{1}{2\Delta\omega\left(\sqrt[3]{\frac{\omega}{\omega_n}} - 1\right)} = \frac{1}{2\Delta\omega\left(\sqrt[3]{\frac{\omega_n + \Delta\omega}{\omega_n}} - 1\right)} \qquad (27)$$

As it is seen, when $\omega = \omega_n$ (or $\Delta\omega = 0$), the time is t→∞, as it should be for a stationary state. For $\Delta\omega \ll \omega_n$, the expression for the time (27) is symmetric (for positive and negative $\Delta\omega$). It is more convenient to transform eqs.(27) in terms of the *binding* energy:

$$t = \frac{\hbar}{2\Delta E\left(\sqrt[3]{\frac{\Delta E}{E_n} + 1} - 1\right)} \qquad (28)$$

where the energy can be measured in units eV and $\hbar$ [eV.s]. In this case the energy of the different excited states can be expressed through the Rydberg constant (R). Thus, the life time of each single excited hydrogen atom depends on the small energy difference ($\Delta E$) and the principal quantum number (*n*):

$$t = \frac{\hbar}{2\Delta E\left(\sqrt[3]{1 + \frac{n^2 \Delta E}{R}} - 1\right)} \qquad (29)$$

In the case when $n^2 \Delta E \ll R$, the cubic root can be expanded in a series, and taking only two first terms of the expansion $(1 + n^2 (\Delta E)/3R ...)$ to give:

$$t = \frac{3\hbar R}{2(\Delta E)^2 n^2} \qquad (30)$$

Part of the results are shown in the Fig.4 (for $\hbar = 6.59 \times 10^{-16}$ eV.s and R=13.595 eV). These curves are different for different excited states (*n*). They could be compared with the normalized "own lifetimes" of nuclei (t/$\tau$ and $\Delta E/\Gamma$) [11].

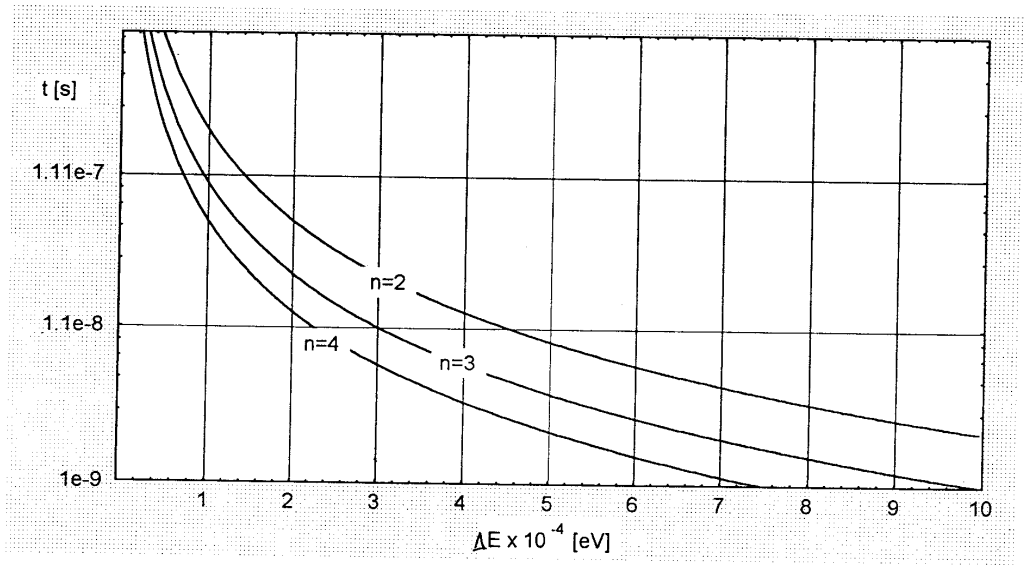

**Fig. 4.** *Time (t) versus energy ($\Delta E$) for n=2,3 and 4. These curves are symmetrical to the curves for energy differences ($-\Delta E$) (to the left of $E_n$).*

## *6. The natural width and mean life time of an ensemble of excited hydrogen atoms*

Similar to the results in [11], the "own life time" (t) of one single excited atom (in state (*n*)) depends exactly on the energy difference ($\Delta E$) (30). The own life time (t [s]) is determined by the exact energy of excitation ($\Delta E = |E_n - E|$), the Planck ($\hbar$) and Rydberg (*R*) constants, and the principle quantum number (*n*) of the excited state. This time cannot be measured experimentally (except in the case shown in [4,11] for resonant Mossbauer transitions in nuclei). Experiments with hydrogen measure only the mean lifetime of an ensemble of excited atoms.

The statistical natural width of the levels ($\Gamma_n$) and mean life times ($\tau_n$) (for different excited states) of an *ensemble* of hydrogen atoms will be found and compared to reference data. Assume that $N_0$ [cm$^{-3}$] atoms (thin target) are irradiated by a flux of photons with uniform energy distribution $\Phi(E) = \Phi_0$[cm$^{-2}$s$^{-1}$] = const.

(in the region of some quantum level *n*). If the effective cross-section of excitation is $\sigma_E$, then, the activity of excited level can be obtained as:

$$\frac{dN}{dt}(t) = \Phi_0 \sigma_E N_0 (1 - \exp(-t/\tau_n)) \qquad (31)$$

As it is well known, after irradiation stops, the activity changes with time in the following way:

$$\frac{dN}{dt}(t) = \Phi_0 \sigma_E N_0 (\exp(-t/\tau_n)) \qquad (32)$$

On the other hand, the differential cross-section ($d\sigma_E$) is:

$$d\sigma_E = \frac{\sigma_0 \Gamma_n dE}{4(\Delta E)^2 + \Gamma_n^2} \qquad (33)$$

($\sigma_0$ is the cross-section in the maximum; $\Gamma_n$ is statistical natural width of level (*n*)). Then the integral cross-section ($\sigma_E$) will be:

$$\sigma_E = \frac{\pi \sigma_0}{2} \qquad (34)$$

Substituting (34) in (32) gives the variation of activity with time after excitation:

$$\frac{dN}{dt}(t) = \Phi_0 \frac{\pi \sigma_0}{2} N_0 (\exp(-t/\tau_n)) \qquad (35)$$

Under the same conditions, but using the differential cross-section (33), shows how activity $\frac{dN}{dt}(E)$ increases with irradiation time:

$$\frac{dN}{dt}(E) = \frac{\Phi_0 N_0 \sigma_0 \Gamma_n dE}{4(\Delta E)^2 + \Gamma_n^2} (1 - \exp(-t/\tau_n)) \qquad (36)$$

To derive an expression for this activity after irradiation, from (30) the variation of the own life time (*t*) with energy is:

$$dt = \frac{3\hbar R dE}{(\Delta E)^3 n^2} \qquad (37)$$

Because of the symmetry of (30), (Figs.5,6) with respect of energy, in the time interval (*dt*) decay the atoms in the two intervals $\Delta E$ on both sides of $E_n$:

$$dt = \frac{3\hbar R dE}{(\Delta E)^3 n^2} + \frac{3\hbar R dE}{(\Delta E)^3 n^2} = \frac{6\hbar R dE}{(\Delta E)^3 n^2} \qquad (38)$$

or

$$dE = \frac{(\Delta E)^3 n^2 dt}{6\hbar R} \qquad (39)$$

Substituting (*dE*) in (36) gives the activity of hydrogen atoms *after irradiation*:

$$\frac{dN}{dt}(E) = \frac{\Phi_0 N_0 \sigma_0 \Gamma_n (\Delta E)^3 n^2 dt}{\left(4(\Delta E)^2 + \Gamma_n^2\right) 6\hbar R} \qquad (40)$$

Two expressions for the activities are found: (40), depends on the energy of excitation ($\Delta E$), and (35), depends on time (t). In the experiments where the energy ($\Delta E$) cannot be measured, the two activities (35) and (40) must be equal [11]:

$$\frac{\Phi_0 N_0 \sigma_0 \Gamma_n (\Delta E)^3 n^2 dt}{\left(4(\Delta E)^2 + \Gamma_n^2\right) 6\hbar R} = \Phi_0 \frac{\pi \sigma_0}{2} N_0 \left(\exp(-t/\tau_n)\right) \qquad (41)$$

In the specific case (Fig.5 and 6) when $\exp(-t/\tau_n) = 1/2$, then $\Delta E = \Gamma_n/2$, and the expression (41) becomes:

$$\frac{\Gamma_n^2 n^2 dt}{24 \hbar R} = \pi \qquad (42)$$

Hence, the natural width ($\Gamma_n$) of a statistical ensemble of atoms (for unit time interval, dt=1) can be calculated as:

$$\Gamma_n = \frac{1}{n} \sqrt{24 \pi \hbar R} \qquad (43)$$

For the population of a statistical ensemble, natural levels width (normalized in the maximum) are shown in Fig. 5.

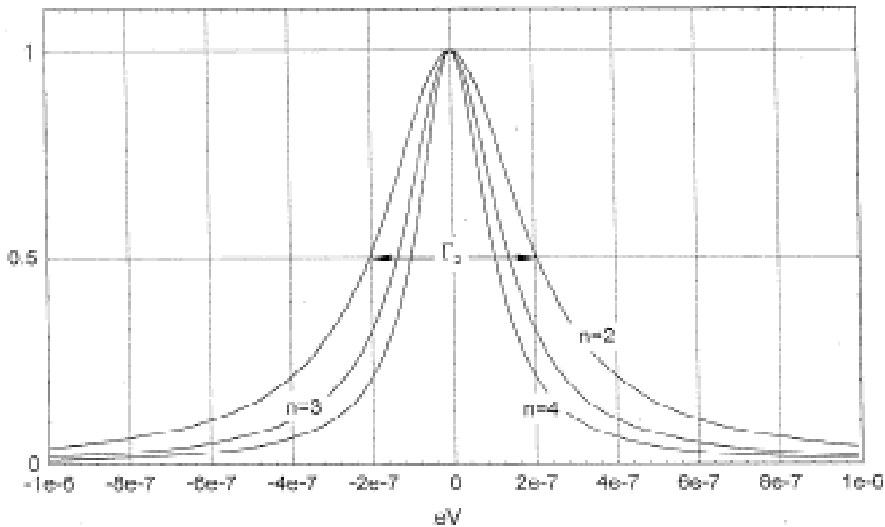

**Fig. 5.** *The normalized natural lines of hydrogen atom (n=2,3 and 4). The energy ($\Delta E = E_n - E$) is calculated in absolute units [eV].*

From the natural width ($\Gamma_n$) of level (*n*) it is easy to derive the mean lifetime of all excited atoms (at level *n*):

$$\tau_n = \frac{\hbar}{\Gamma_n} = n\sqrt{\frac{\hbar}{24\pi R}} \qquad (44)$$

Thus, for calculation of the mean life time of an statistical ensemble of excited hydrogen level (*n*), only Rydberg's constant (*R*) and Planck's constant ($\hbar$) are needed. The corresponding decay constant (the spontaneous coefficient of Einstein) is $A_n = 1/\tau_n$.

Fig. 6. Time ($t/\tau$) versus energy ($\Delta E / \Gamma_n$) of excitations and normalized natural width of the first excited state. When decay moves from the wings of the level ($|\Delta E| = \infty$) to the place $\Delta E = \Gamma_n / 2$, then half of the excited atoms have decayed and $\exp(-t/\tau_n) = 1/2$.

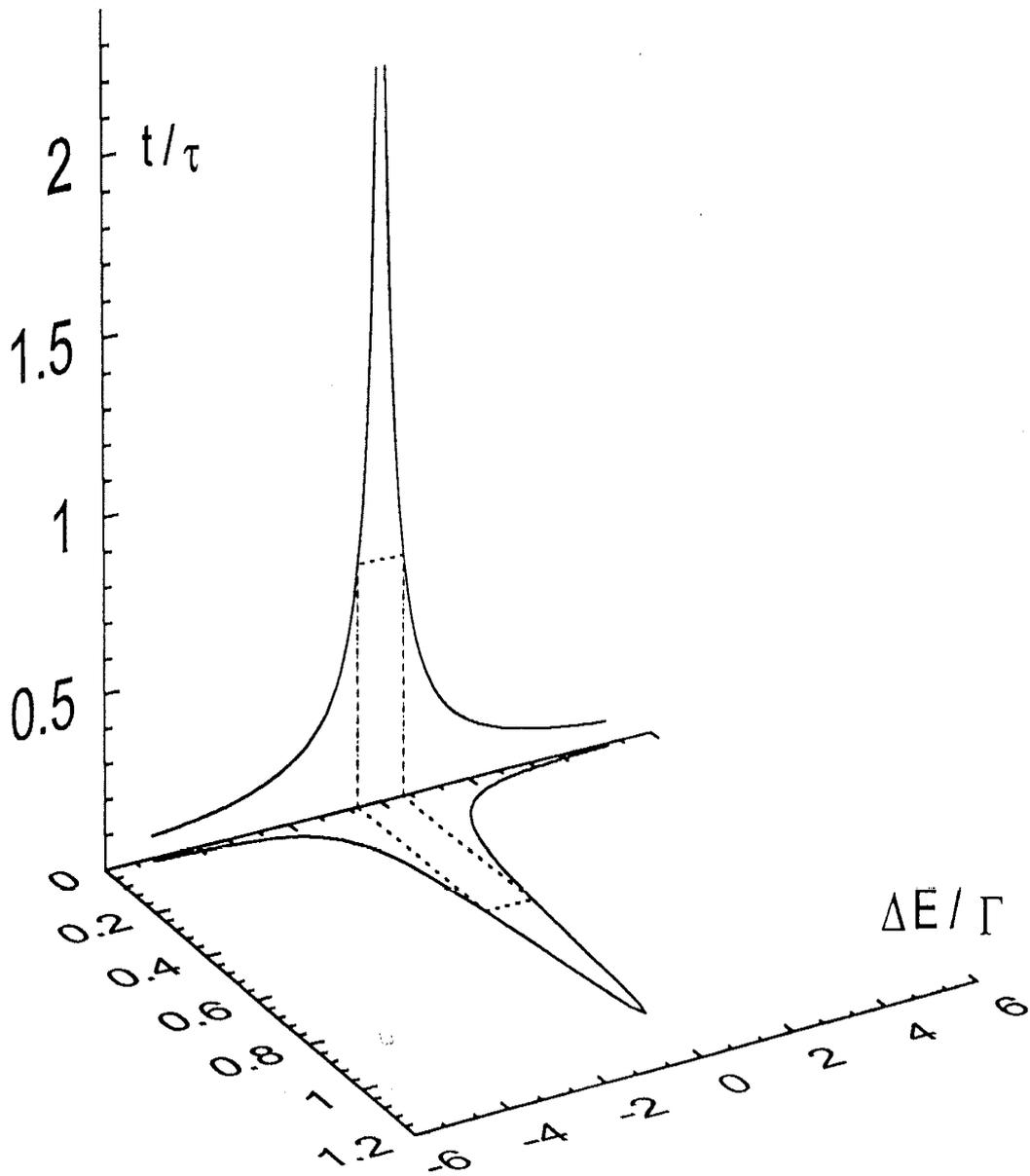

**Fig. 6.** *Time (t/$\tau$) versus energy ($\Delta E / \Gamma_n$) of excitations and normalized natural width of the first excited state. When decay moves from the wings of the level ($|\Delta E| = \infty$) to the place $\Delta E = \Gamma_n / 2$, then half of the excited atoms have decayed and $\exp(-t / \tau_n) = 1/2$.*

## 7. Comparison with reference data.

In the numerous reference tables on hydrogen gave quite different values for $\tau_n$ (especially for low binding energy of the excited states; $n > 2$). In Table 1 below there are data from [12] (1966) and [13] (1986) compared to the calculations (formula 44, 1997) [4,5].

**Table 1.** *The values of $\tau_n = 1/A_n$ (and natural width of the levels) from present paper (1997) are closer to the values of data source [13] (1986). The difference between the data from [12] (1966) and [13] (for n>2) are impermissible.*

| | Data Sources | | | | | |
|---|---|---|---|---|---|---|
| | [12] (1966) | | [13] (1986) | | (1997) | |
| n | $\tau_n$, s | $A_n$, s$^{-1}$ | $\tau_n$, s | $A_n$, s$^{-1}$ | $\tau_n$, s | $A_n$, s$^{-1}$ |
| 2 | $2.12 \times 10^{-9}$ | $4.699 \times 10^8$ | $1.60 \times 10^{-9}$ | $6.25 \times 10^8$ | $1.603 \times 10^{-9}$ | $6.23 \times 10^8$ |
| 3 | $1.0 \times 10^{-8}$ | $1.0 \times 10^8$ | $3.94 \times 10^{-9}$ | $2.53 \times 10^8$ | $2.405 \times 10^{-9}$ | $4.15 \times 10^8$ |
| 4 | $3.3 \times 10^{-8}$ | $3.02 \times 10^7$ | $8.0 \times 10^{-9}$ | $1.24 \times 10^8$ | $3.2 \times 10^{-9}$ | $3.12 \times 10^8$ |

As it is seen, for the second excited state (*n*=2) the calculated $\tau_n$ is equal to $1.603 \times 10^{-9}$ s, while in [12] this time is $\tau_n = 2.127 \times 10^{-9}$ s and in [13] $\tau_n = 1.60 \times 10^{-9}$ s. So, the result from the present calculations is in excellent agreement with reference data [13] (for *n*=2). It is necessary to stress that the calculations fit better to the values in [13] (1986). The differences between the values in [12] and [13] are greater than the differences between the calculations and the data in [13]. So, the *Bohr's model (complemented with de Broglie's ideas) continues to describe hydrogen properties (mean life time, natural width of the levels) as exactly as Bohr's hydrogen model describes the frequency of radiation.*

## *8. Differences between the data.*

As it is known, the experimental accuracy for frequency measurement is very great in comparison with accuracy of time measurements. An attempt to explain the great differences [12,13] between reference data (for *n*>2) will be made. Experimental results are very good only for the first excited states... The differences between reference data (for $n > 2$) are caused by experimental difficulties and incorrect application of the relation between Einstein's coefficients, which is explained in [14,15].

In [12] the transition probability for spontaneous emission from upper state k to lower state i, $A_{ki}$, is related to the total intensity $I_{ki}$ of a line of frequency $\nu_{ik}$ by

$$I_{ki} = \frac{1}{4\pi} A_{ki} h\nu_{ik} N_k \quad \text{(expression (1) on page ii of [12])} \tag{45}$$

where $h$ is Planck's constant and $N_k$ the population of state $k$. It was shown in [14,15] that this relation holds for transitions from any excited state $k$ to the ground state i only. If (i) is also an excited state, then relation (45) must be:

$$I_{ki} = \frac{1}{4\pi}(A_{ki} + \frac{g_i}{g_k} A_{ix}) h\nu_{ik} N_k \tag{46}$$

where $A_{ix}$ is the full decay constant of level (i) and $g_i$, $g_k$ are the corresponding statistical weights. Only when $A_{ix}=0$ (ground state), (46) coincides with (45). The same applies for the transition probability of absorption $B_{ik}$ and the transition probability of induced emission $B_{ki}$ in [12]:

$$B_{ik} = 6.01 \lambda^3 \frac{g_k}{g_i} A_{ki} \quad \text{(expr. (6), p. vi of [12])} \tag{47}$$

$$B_{ki} = 6.01 \lambda^3 A_{ki} \quad \text{(expr. (7), p. vi of [12])} \tag{48}$$

($\lambda$ is the wavelength in Angstrom units). When (i) is an excited state, these relations are also wrong. According to [14,15], these relations (in the same units as in [12]) will be:

$$B_{ik} = 6.01 \lambda^3 \left( \frac{g_k}{g_i} A_{ki} + A_{ix} \right) \tag{49}$$

$$B_{ki} = 6.01 \lambda^3 \left( A_{ki} + \frac{g_i}{g_k} A_{ix} \right) \tag{50}$$

It is also seen that if (i) is a ground state, $A_{ix} = 0$, these relations correspond to the relations in [12]. It is clear that even based on experimental results (when $n > 2$), $\tau_n$ can have wrong values if processed using the inappropriate (but commonly accepted) relations [12].

*The mean lifetimes of excited levels of the simplest atom - hydrogen - obtained herein are in surprising agreement with the known data. At the same time, the differences between the reference values for n>2, shows that all reference data for transition probabilities in hydrogen must be critically examined and adjusted accurately according the present results.*

## 9. The shape of acceleration curve and the shape of electromagnetic field of the soliton.

On the Fig.2 it is shown only one example of the acceleration curve. This curve cannot be known exactly, because the shape of electromagnetic field of the soliton (Fig.1) also is not known exactly [3]. We know from electrodynamics only that the shape of two curves must exactly coincide. The two curves can be symmetrical or not symmetrical, but independently of its exact shape we can calculate the integral values of necessary parameters for solitons (acceleration) emitted by hydrogen atom (8),(10). The effective time of transition (acceleration) is

$$t_e = \frac{1}{2\omega} = \frac{\hbar^3}{e^4 m_0 \left(\frac{1}{n^2} + \frac{1}{k^2}\right)} \qquad (51)$$

Effective acceleration (*a*) is:

$$a = \frac{e^6 m_0}{\hbar^4}\left[\left(\frac{1}{n} - \frac{1}{k}\right)\left(\frac{1}{n^2} - \frac{1}{k^2}\right)\right] \qquad (52)$$

and the effective path $H_{nk}$ is:

$$H_{nk} = \frac{\hbar^2}{e^2 m_0}\left[\frac{1}{2\left(\frac{1}{n}+\frac{1}{k}\right)} + \frac{1}{k\left(\frac{1}{n^2} - \frac{1}{k^2}\right)}\right] \qquad (53)$$

As it is seen for $n = 1$ and $k = 2$ the effective path (53) is equal to the Bohr's radius $r_0$:

$$H_{nk} = \frac{\hbar^2}{e^2 m_0} = r_0 \approx 5.3 \times 10^{-11} \text{m} \qquad (54)$$

and it is smaller from the distance between the two orbits ($3 r_0$). The energy of the soliton $E_{nk}$ (12), as it must be, do not depend on effective time $t_e$ (51), but this effective time is necessary for comparison between the shapes of two curves: electric field of soliton and acceleration of electron.

**The possible shape of electromagnetic soliton.**

The most often the shape of soliton curve (with a form like a bell) is described from equation [16]:

$$E = (2/\tau)\text{sech}[(t - x/v)/\tau] \qquad (55)$$

Where $\tau$ is connected with the width of the impulse. (55) can be only one example; it is not sure that electromagnetic field of the soliton [3] correspond exactly to (55), but according to definition in [3] the electric field in the maximum of the curve ($E_0$) and effective time ($t_e$) are related:

$$E_0 t_e = \int_{-\infty}^{\infty} E(t)dt \qquad (56)$$

The electric field of the soliton in vacuum can be written:

$$E(t) = E_0 \frac{2}{(e^{t/t_e} + e^{-t/t_e})} \quad \text{or} \quad E(x) = E_0 \frac{2}{(e^{x/l_e} + e^{-x/l_e})} \qquad (57)$$

If the soliton electric field correspond to (57), then acceleration of electron ($a(x)$) must have the same shape (in the space along the unknown trajectory $x$, but known effective path $H_{nk}$):

$$a(x) = a_0 \frac{2}{(\exp(x/H_{nk}) + \exp(-x/H_{nk}))} \qquad (58)$$

Here $a_0$ is acceleration in the maximum ($a_0 \sim E_0$) and

$$a_0 H_{nk} = \int_{-\infty}^{\infty} a(x)dx \qquad (59)$$

The shapes of the two curves (57 and 58) cannot be accepted as exact, but it is sure that they must correspond to each other and the equations (56 and 59) are exact by definition. Knowing effective length of electromagnetic soliton in vacuum ($l_e$) and effective path of electron ($H_{nk}$) one can estimate the average velocity of electron ($v_{nk}$) when passing the distance $H_{nk}$:

$$H_{nk} = v_{nk} t_e \qquad (60)$$

and

$$l_e = ct_e \tag{61}$$

So, the ratio of the two velocities is

$$v_{nk}/c = H_{nk}/l_e \tag{62}$$

When the transition (hydrogen) occur between $k=2$ and $n=1$ this ratio is:

$$v_{nk}/c \approx 5.6 \times 10^{-3}$$

The velocity of the electron is about 3 orders of magnitude smaller than the velocity of the light.

## 10. *Some inevitable conclusions.*

Here we are finding the effective acceleration of electron in hydrogen but it is evident that the acceleration vector changes in time and direction. These details I cannot find now because *only integral* acceleration is known. If we know the exact shape of electromagnetic field of the solitons (Fig.1), the exact shape of acceleration curve (Fig.2) would be known also and vise versa. Now we can make only some supposition about the changes of acceleration vector in space and time. This vector is probably not radial, as it is not radial the path ($H_{nk}$) of the electron. The acceleration vector describes some complicated curve different of acceleration in an ordinary dipole and it is sure that emission of a hydrogen atom cannot be presented as emission of a vibrating dipole. In the beginning and on the end the acceleration of electron must be zero and must have one positive maximum (Fig.1 and Fig.2). The acceleration vector must describes some curve in the space for a very short effective time $t_e$. This curve probably lay on the plane of velocity vectors and determines the direction of soliton propagation.

Using only the bases of classical electrodynamics we can conclude that electromagnetic field in every moment is emitted perpendicularly to the acceleration vector in this moment. Electromagnetic field must be redistributed in different directions and cannot form a soliton (electromagnetic particle) which propagate only in one direction.

These simple calculations show that the most difficult question about the time of transition in atoms can be answered combining the soliton properties and the Bohr's model of atom. We think some inevitable changes in classical electrodynamics are needed and this probably will change the quantum electrodynamics also.

Now it is clear that energy of the soliton is discrete in time and space as it is discrete acceleration of the electron. We else do not know what is the exact shape of acceleration and the exact shape of soliton electromagnetic field (Fig.1 and Fig.2), but I hope that they can be found and consequently it can be found the direction of the soliton propagation.

The calculations show that Bohr's model of hydrogen is as useful as the real field of L. de Broglie is. *The hydrogen atomic model of Bohr-de Broglie allow (for the first time) to calculate exactly the mean lifetime ($\tau_n = 1/A_n$) of an ensemble of the excited levels*. If the energy of excitation is different from that corresponding to the exact conditions for a stationary state, after some evolution of the excited state the Coulomb field can change the state of the electron (acceleration) because the amplitude of de Broglie's waves becomes zero and the electron is no more in a potential well (the electron can emit a photon-soliton [1,2,3]). The main result of this work is that excited states of the hydrogen atom decay after some exactly predictable time (t) (30) and the emission of the photon-soliton takes an exact time ($\sim t_e$). Decay is not an accidental event as it is believed by the majority of scientists (except Einstein who wrote that a weakness of the theory of radiation is that the time of occurrence of an elementary process is left to "chance"). *The mean life time ($\tau_n$) is a characteristic only of a statistical ensemble of excited atoms* (40).

If a transition occurs between two excited states ($E_n = R(1/n^2)$ and $E_k = R(1/k^2)$) the frequency of the emitted soliton is calculated according to: $(E_n - E_k)/\hbar = \omega_{nk}$. This frequency correspond to the maximum of the line distribution of a statistical ensemble of hydrogen atoms. The width of the photon line is the sum of the levels width: $\Gamma_{nk} = \Gamma_n + \Gamma_k$.

So, for a *statistical ensemble* of hydrogen atoms the distances of electrons from the protons (or energies), are very different. In such an ensemble the probability to find an electron at some distance (or energy) from the proton is maximal at the places of Bohr's stationary orbits. This probability is smaller at other places, but never becomes zero. For the coordinate systems related with the *center of mass of every hydrogen atom*, these probabilities are presented on Fig.5. For a laboratory coordinate system the probabilities depend by motions of the center of masses and become consistent with contemporary quantum physics.

I think that all these solitary objects do not contradict Quantum Mechanics (QM) - especially the properties of a solitary hydrogen atom - but only reveal some unknown details of SQS. *It seems to me now, that the properties of a solitary quantum object must be different from the properties of the statistical ensemble of such objects and cannot be further neglected...* As it is known, Bohr's model of the hydrogen atom concerns a *solitary hydrogen atom*. However, all experimental results which confirm this model - excited states, frequency of the emitted lines, the calculation of Rydberg's constant and so on, are obtained from spectroscopic data about statistical ensembles of hydrogen atoms. The frequencies, exactly calculated by Bohr, correspond to the frequencies of the stationary states (at the maxima of the lines). Now, it is clear that the natural width of the lines (statistical ensemble) cannot be obtained from Bohr's conditions only. Bohr himself probably could solve this problem, if he had accepted de Broglie's ideas about the coexistence of waves and particles.

To pay honor to Luis de Broglie who wrote:
"In the spring of my life, I was obsessed with the problems of quanta and the coexistence of waves and particles in the world of micro-physics: I made decisive efforts, although incomplete, to discover the solution. Now, in the autumn of my existence, the same problem still preoccupies me because, despite of the many successes and the long way passed, I do not believe that the enigma is indeed resolved. The future, a future which I undoubtedly will never see, will maybe decide the problem: it will tell whether my present point of view is an error of an already sufficiently old man who is still devoted to the ideas of his youth, or, on the contrary, this is a clairvoyance of a researcher who all his life has meditated on the most important question of contemporary Physics". (L. de Broglie, *Certitudes et incertitudes de la Science*, Edition Albin Michel, Paris, 1966, p. 22; a free translation from French).

I realize that there are not some consecutive descriptions of a new electrodynamics of electromagnetic fields and charged particles. The known properties of solitons [3] and hydrogen atom [4,5] permit to think about and to describe some fragmentary, *but very convincing and essential results,* concerning the reality of solitons and its interactions with charged particles. It is evident that this fragmentary electrodynamics is not complete, but I hope that in the future it can be completed.

**References:**

1. J. P. Vigier, Foundations of Physics, 21,2, 125 (1991).

2. T. Waite, Ann. Fond. Louis de Broglie (Paris) 20, 4, (1995) 427

3. P. Kamenov, B. Slavov, Foundations of Physics Letters, v.11, No4, (1998) 325.

4. P. S. Kamenov, *Physics of solitary quantum systems*, Paradigma, Sofia, 1999, p.56 – 74;

5. P. Kamenov, Comp. Rend. Acad. Bulg. Sci., 52, 5/6, (1999) 27

6. A. Einstein, Mitt. Phys. Ges. (Zuerich), 18, (1916), p. 47-62

7. A. Einstein, Verh. Dtsch. Phys. Ges., 18 (1916) p.318

8. P. Kamenov, I. Christoskov, Phys.Lett. A, v.140, 1,2 (1989) 13-18.

9. L. de Broglie, Rev. Sci. Prog. Decouvert, 3432, (April 1971), 44

10. M. Born, Z. Phys. 37 (1926) 863; - M. Born, Z. Phys. 38 (1926) 803.

11. P. Kamenov, Nature Phys.Sci. v.231 (1971) 105-107.

12. *Atomic Transition Probabilities*, Volume I *Hydrogen Through Neon* (A critical Data Compilation), W. L. Wiese, M. W. Smith, and B. M. Glennon, National Bureau of Standards, (May 20, 1966).

13. A. A. Radzig, B. M. Smirnov, *Parametry atomov i atomnih ionov* (Data), ENERGOATOMIZDAT, Moskva, 1986 (in Russian).

14. P. Kamenov, A. Petrakiev, and A. Apostolov, Laser Physics, Vol.5, No 2, (1995) 314-317.

15. P. Kamenov, Nuovo Cimento D, 13/11, Nov. 1991, 1369-1377

16. R. K. Bullough, P. J. Caudrey, "Solitons", Ed. Springer – Verlag Berlin Heidelberg New York 1980 (Russian Ed. Moskwa, "Mir" 1983 p. 85).